\documentstyle[12pt,aps]{revtex}
\newcommand{\bold}[1]{\mbox{\boldmath $#1$}}
\tighten
\begin{document}

\title{\flushleft Potentials of a uniformly moving point charge in
the Coulomb gauge\footnote{This paper is written by V Hnizdo in his
private capacity. No official support or endorsement by the Centers for
Disease Control and Prevention is intended or should be inferred.}} 
\author{\flushleft V Hnizdo}
\address{\flushleft National Institute for Occupational Safety and Health,
Morgantown, West Virginia 26505, USA
\newline
\newline
E-mail: vbh5@cdc.gov}
\maketitle

\begin{abstract}
\flushleft
{\bf Abstract}\\
The Coulomb-gauge vector potential of a uniformly moving
point charge is obtained by calculating
the gauge function for the transformation between the Lorenz
and Coulomb gauges. 
The expression obtained for the difference between the
vector potentials in the two gauges is shown to satisfy
a Poisson equation to which the inhomogeneous wave equation
for this quantity can be reduced.
The right-hand side of the Poisson equation
involves an important but easily overlooked delta-function term 
that arises from a second-order partial derivative of the Coulomb potential 
of a point charge.
\end{abstract}
\vspace{5ex}
\begin{flushleft}
{\bf 1. Introduction}
\end{flushleft}
\noindent
Gauge invariance is an important property
of electrodynamics. Notwithstanding the thorough attention it has received
in textbooks \cite{Jackson,Griffiths,PP,LL}, the topic of gauge
invariance seems to be still very much alive.
Recently, articles have appeared on the resolution of apparent
causality problems in the Coulomb gauge \cite{Rohrlich},
the transformation from the Lorenz gauge to the Coulomb and some other
gauges \cite{Jack}, the Coulomb-gauge vector potential in terms of 
the magnetic field \cite{Stew},
and  the historical development of the whole concept of gauge invariance 
\cite{JackOk}.
A discordant voice in this is a claim by Onoochin \cite{Ono} that the electric 
field of a uniformly moving point charge comes out differently when 
it is calculated in the Lorenz and Coulomb gauges, which that author
takes as evidence that the two gauges are not physically equivalent.

Curiously, it is difficult to find in the literature an explicit expression 
for the Coulomb-gauge vector potential of a uniformly moving charge. 
To the present author's knowledge, such an expression has appeared only
in a short comment \cite{Lab},  where it is obtained directly
from the well-known electric field and Coulomb-gauge scalar potential
of the charge. 
The formulae of Jackson \cite{Jack} for the transformation between the 
Lorenz and Coulomb gauges thus have come timely
to provide an analytical check on the claim of Onoochin.
The purpose of the present paper is to clear up the problem
of the Coulomb-gauge vector potential of a uniformly moving charge     
using Jackson's results as well as the method employed by Onoochin,
in which the difference between the Coulomb- and Lorenz-gauge vector potentials 
in this problem is to be found by solving a Poisson equation.

In section 2, we obtain the difference 
between the potentials in the Lorenz and Coulomb gauges
for a uniformly moving point charge using the formalism of Jackson's
paper \cite{Jack},
which guarantees that the two gauges will yield the same electric and
magnetic fields. In section 3, we demonstrate that the expression obtained for 
the difference  between the vector potentials in the two
gauges satisfies the inhomogeneous wave equation for this difference,
and in section 4 we show where and why the procedure used by Onoochin
for solving that wave equation by reducing it to a Poisson equation
went wrong. It will turn out that the right-hand side of this
equation must include an easily overlooked delta-function term to yield 
the correct solution. Concluding remarks are made
in section 5, and an appendix contains some calculational details.

\begin{flushleft}
 {\bf 2. Transformation from the Lorenz gauge to the Coulomb gauge}
\end{flushleft}
\noindent
The scalar potential $V$ and the vector potential $\bold{ A}$  of a uniformly 
moving point charge in the Lorenz gauge, defined by the condition
$\bold{\nabla\cdot}\bold{ A}+\partial\bold{ A}/c\partial t =0$,
are well known. For a point charge $q$
moving with a constant velocity $\bold{ v}=v \hat{\bold{x}}$, the Lorenz-gauge 
potentials $V_{\rm L}$ and $\bold{ A}_{\rm L}$ are given by (we shall use the Gaussian 
system of units):
\begin{equation}
V_{\rm L}(\bold{ r},t)=\frac{q}{\sqrt{(x-v t)^2+\frac{1}{\gamma^2}(y^2+z^2)}}
\;\;\;\;\;\;
\bold{ A}_{\rm L}(\bold{ r},t)=\frac{\bold{ v}}{c}V_{\rm L}(\bold{ r},t)
\label{VLAL}
\end{equation}
where $\gamma=(1-v^2/c^2)^{-1/2}$ and, for simplicity, the charge is assumed 
to pass through the origin $x{=}y{=}z{=}0$ at time $t=0$ 
(see, e.g., \cite{PP}, section 19-3).
In the Coulomb gauge, defined by the condition $\bold{\nabla\cdot}\bold{ A}=0$,
the scalar potential $V_{\rm C}$ of the charge takes a particularly simple form,
\begin{equation}
V_{\rm C}(\bold{ r},t)=\frac{q}{\sqrt{(x-v t)^2+y^2+z^2}}
\label{VC}
\end{equation}
as the scalar potential in this gauge is exactly the same as that
of the instantaneous Coulomb interaction of electrostatics. 
On the other hand, the Coulomb-gauge vector potential $\bold{ A}_{\rm C}$
is the retarded solution to a relatively complicated inhomogeneous
wave equation 
\begin{equation}
\Box \bold{ A}_{\rm C}
=-\frac{4\pi}{c}q\bold{ v}\delta(\bold{ r}-\bold{ v}t)
+\bold{\nabla}\frac{\partial V_{\rm C}}{c\partial t}
\label{Cweq}
\end{equation}
where $\Box=\nabla^2-\partial^2/c^2\partial t^2$ is the d'Alembertian
operator and
$q\bold{ v}\delta(\bold{ r}-\bold{ v}t)$ is the point-charge
current density of the present problem (see, e.g., \cite{Jackson}, section 6.3).

The gauge invariance of electrodynamics implies that there is
a gauge function $\chi_{\rm C}$ that 
connects the Coulomb- and Lorenz-gauge potentials by
\begin{equation}
V_{\rm C}=V_{\rm L}-\frac{\partial\chi_{\rm C}}{c\partial t}\;\;\;\;\;\;
\bold{ A}_{\rm C}=\bold{ A}_{\rm L}+\bold{\nabla}\chi_{\rm C}
\label{connect}
\end{equation}
which ensures that the Lorenz-gauge and Coulomb-gauge potentials will yield
the same electric and magnetic fields. This is because the fields
are generated from the potentials via the prescription
\begin{equation}
\bold{ E}(\bold{ r},t)=-\bold{\nabla}V(\bold{ r},t)
-\frac{\partial\bold{ A}(\bold{ r},t)}{c\partial t}\;\;\;\;\;\;
\bold{ B}(\bold{ r},t)=\bold{\nabla\times}\bold{ A}(\bold{ r},t)
\label{prescription}
\end{equation}
and thus any electric-field and magnetic-field differences
that could arise from the use of the different gauges are guaranteed
to vanish:
\begin{eqnarray}
-\bold{\nabla}(V_{\rm C}-V_{\rm L})-\frac{\partial(\bold{ A}_{\rm C}-\bold{ A}_{\rm L})}{c\partial t}
&=&\bold{\nabla}\frac{\partial\chi_{\rm C}}{c\partial t}
-\frac{\partial}{c\partial t}
\bold{\nabla}\chi_{\rm C}= 0\\
\bold{\nabla\times}(\bold{ A}_{\rm C}-\bold{ A}_{\rm L})
&=&\bold{\nabla\times\nabla}\chi_{\rm C}= 0.
\end{eqnarray}
Nevertheless, it should be instructive to demonstrate explicitly 
that it is indeed the case also in the present problem by finding the requisite
gauge function. Before we turn to this task,
we give here for completeness the fields that the prescription
(\ref{prescription}) yields with
the Lorenz-gauge potentials $V_{\rm L}$ and $\bold{ A}_{\rm L}$ of equation
 (\ref{VLAL}):
\begin{equation}
\bold{ E}(\bold{ r},t)=q\frac{\gamma(\bold{ r}-\bold{ v}t)}
{[\gamma^2(x-v t)^2+y^2+z^2]^{3/2}}
\;\;\;\;\;\;
\bold{ B}(\bold{ r},t)=\frac{1}{c}\bold{ v}\bold{\times}\bold{ E}(\bold{ r},t).
\end{equation}
The same expressions for the electric and magnetic fields 
can be obtained also by
Lorentz-transforming the electrostatic Coulomb field of the charge
from its rest frame to the `laboratory' frame.

Jackson \cite{Jack} has derived the following integral expression for 
the gauge function $\chi_{\rm C}$ in terms of the charge density:
\begin{equation}
\chi_{\rm C}(\bold{ r},t)=-c\int {\rm d}^3r'\frac{1}{R}\int_0^{R/c}{\rm d}
\tau\rho(\bold{ r}',t-\tau)
\label{chic}
\end{equation}
where $R=|\bold{ r}-\bold{ r}'|$ (a gauge function  is defined
to within an arbitrary additive constant, which we omit here).
For a point charge $q$ moving with a constant velocity $\bold{v}$
along the $x$-axis, the charge density is
\begin{equation}
\rho(\bold{ r},t)=q\delta(\bold{ r}-\bold{ v}t)=q\delta(x-vt)\delta(y)\delta(z). 
\label{rho}
\end{equation}
This gives
\begin{eqnarray}
\int_0^{R/c}{\rm d}\tau\rho(\bold{ r}',t-\tau)
&=&q\delta(y')\delta(z')\int_0^{R/c}{\rm d}\tau\,
\delta[x'-v(t-\tau)]
\nonumber\\
&=&\frac{q}{|v|}\delta(y')\delta(z')\int_0^{R/c}{\rm d}\tau\,
\delta[\tau-(t-x'/v)]\nonumber \\
&=&\frac{q}{|v|}\delta(y')\delta(z')\{\Theta[R/c-(t-x'/v)]
-\Theta[-(t-x'/v)]\}\nonumber \\
&=&\frac{q}{v}\delta(y')\delta(z')[\Theta(x'-x_0)-\Theta(x'-vt)]
\end{eqnarray}
where $\Theta(x)$ is the Heaviside step function and
\begin{equation}
x_0=x-\gamma^2\left[x-vt+\frac{v}{c}
\sqrt{(x-vt)^2+\frac{1}{\gamma^2}(y^2+z^2)}\right]\;\;\;\;\;
\gamma=\frac{1}{\sqrt{1-v^2/c^2}}.
\label{x0}
\end{equation}
The gauge function (\ref{chic}) with the charge density (\ref{rho}) is thus
\begin{eqnarray}
\chi_{\rm C}(\bold{ r},t)&=&-q\frac{c}{v}\int_{-\infty}^{\infty}{\rm d}x'\,
\frac{\Theta(x'-x_0)-\Theta(x'-vt)}{\sqrt{(x-x')^2+y^2+z^2}}
\nonumber \\
&=&-q\frac{c}{v}\int_{x_0}^{vt}\frac{{\rm d}x'}{\sqrt{(x-x')^2+y^2+z^2}}
\nonumber \\
&=&q\frac{c}{v}\left[
{\rm arcsinh}\frac{x-vt}{\sqrt{y^2+z^2}}
-{\rm arcsinh}\frac{x-x_0}{\sqrt{y^2+z^2}}\right].
\label{chic2}
\end{eqnarray}

Let us first check that the gauge function (\ref{chic2})
yields the established difference $V_{\rm C}-V_{\rm L}$ between the scalar 
potentials in the two gauges.
Using the identity
\begin{equation}
\frac{1}{\gamma^2}\sqrt{(x-x_0)^2+y^2+z^2}
=\frac{v}{c}(x-v t)+\sqrt{(x-vt)^2+\frac{1}{\gamma^2}(y^2+z^2)}
\end{equation}
which follows from (\ref{x0}), to simplify the result of the partial
differentiation $-\partial\chi_{\rm C}/c\partial t$, we obtain
\begin{equation}
V_{\rm C}-V_{\rm L}=-\frac{\partial\chi_{\rm C}}{c\partial t}
=\frac{q}{\sqrt{(x-vt)^2+y^2+z^2}}
-\frac{q}{\sqrt{(x-vt)^2+\frac{1}{\gamma^2}(y^2+z^2)}}.
\label{difV}
\end{equation}
This is indeed the correct result [see equations (\ref{VLAL}) and (\ref{VC})].

Calculating the $x$-component of the difference $\bold{ A}_{\rm C}-\bold{ A}_{\rm L}$
between the vector potentials in the two gauges is now very simple
because $\partial\chi_{\rm C}/\partial x=-(1/v)\partial\chi_{\rm C}/\partial t$
on account of the dependence of the gauge function (\ref{chic2}) on
the variables $x$ and $t$ only through the combination $x-vt$.
We thus have
\begin{equation}
A_{{\rm C}\,x}- A_{{\rm L}\,x}=\frac{\partial\chi_{\rm C}}{\partial x}
=-\frac{c}{v}\frac{\partial\chi_{\rm C}}{c\partial t}=\frac{c}{v}(V_{\rm C}-V_{\rm L}).
\label{ACx}
\end{equation}
The $y$- and $z$-components of the difference
$\bold{ A}_{\rm C}-\bold{ A}_{\rm L}=\bold{\nabla}\chi_{\rm C}$
are obtained by performing direct differentiations in a similar way to
that of calculating the value (\ref{difV}) for the difference $V_{\rm C}-V_{\rm L}$,
yielding
\begin{eqnarray}
A_{{\rm C}\,y}-A_{{\rm L}\,y}&=&\frac{\partial\chi_{\rm C}}{\partial y}
=-\frac{c}{v}\frac{y(x-vt)}{y^2+z^2}(V_{\rm C}-V_{\rm L})
\label{ACy}\\
A_{{\rm C}\,z}-A_{{\rm L}\,z}&=&\frac{\partial\chi_{\rm C}}{\partial z}
=-\frac{c}{v}\frac{z(x-vt)}{y^2+z^2}(V_{\rm C}-V_{\rm L}).
\label{ACz}
\end{eqnarray}
These components have no singularities (they vanish at $x-vt=y=z=0$).
As $A_{{\rm L}\,y}=A_{{\rm L}\,z}=0$, equations (\ref{ACy}) and (\ref{ACz}) also
give the Coulomb-gauge components $A_{{\rm C}\,y}$ and $A_{{\rm C}\,z}$
themselves, respectively.

It is instructive to perform the Lorentz transformation of the
Coulomb-gauge four-potential $(V_{\rm C},\bold{ A}_{\rm C})$ from the `laboratory'
frame, where the charge moves with the constant velocity 
$\bold{ v}=v\hat{\bold{ x}}$,
to its rest frame (the primes denote the rest-frame quantities):
\begin{eqnarray}
V_{\rm C}'&=&\gamma(V_{\rm C}-vA_{{\rm C}\,x}/c)\label{Ltr}\\
A_{{\rm C}\,x}'&=&\gamma(A_{{\rm C}\,x}-vV_{\rm C}/c)\;\;\;\;\;\;A_{{\rm C}\,y}'=A_{{\rm C}\,y}
\;\;\;\;\;\;A_{{\rm C}\,z}'=A_{{\rm C}\,z}.
\end{eqnarray}
We note first that with the $x$-component $A_{{\rm C}\,x}$ of the vector potential
in the Coulomb gauge given by
\begin{equation}
A_{{\rm C}\,x}=\frac{c}{v}(V_{\rm C}-V_{\rm L}/\gamma^2)
\end{equation}
which follows from the expression (\ref{ACx}) for the difference 
$A_{{\rm C}\,x}-A_{{\rm L}\,x}$ and the fact that $A_{{\rm L}\,x}=(v/c)V_{\rm L}$,
equation (\ref{Ltr}) gives the rest-frame scalar potential $V_{\rm C}'$ as
\begin{equation}
V_{\rm C}'=\frac{1}{\gamma}V_{\rm L}
=\frac{q}{\gamma\sqrt{(x-vt)^2+(y^2+z^2)/\gamma^2}}
=\frac{q}{\sqrt{{x'}^2+{y'}^2+{z'}^2}}
\end{equation}
where the Lorentz transformation of the coordinates, $x'=\gamma(x-vt)$,
$y'=y$, $z'=z$, is performed on the right-hand side. The rest-frame
scalar potential $V_{\rm C}'$ is simply that of a point charge $q$
in electrostatics. We note also that because the Coulomb-gauge condition
$\bold{\nabla\cdot}\bold{ A}=0$
is not Lorentz invariant, the rest-frame vector potential $\bold{ A}_{\rm C}'$
is not divergenceless in the rest-frame variables $x'$, $y'$, $z'$; 
the potentials $V_{\rm C}'$, $\bold{ A}_{\rm C}'$ are therefore no longer those
of the Coulomb gauge. However, a direct calculation shows that 
the vector potential $\bold{ A}_{\rm C}'$
is irrotational, $\bold{\nabla}'\bold{\times}\bold{ A}_{\rm C}'=0$, which expresses
the fact that there is no magnetic field in the rest frame. Moreover,
the vector potential $\bold{ A}_{\rm C}'$ is independent of the rest-frame time $t'$,
and thus the electric field in the rest frame is given only by
$\bold{ E}'=-\bold{\nabla}'V_{\rm C}'$, which yields correctly the electrostatic
Coulomb field of a charge at rest.

\begin{flushleft} 
{\bf 3. Inhomogeneous wave equation for the vector-potential difference}
\end{flushleft}
\noindent
The difference $\bold{ A}_{\rm C}-\bold{ A}_{\rm L}$ must satisfy the inhomogeneous
wave equation
\begin{equation}
\Box (\bold{ A}_{\rm C}-\bold{ A}_{\rm L})
=\bold{\nabla}\frac{\partial V_{\rm C}}{c\partial t}
\label{(C-L)weq}
\end{equation}
which is obtained by subtracting the wave equation for $\bold{ A}_{\rm L}$,
\begin{equation}
\Box \bold{ A}_{\rm L}=-\frac{4\pi}{c}q\bold{ v}\delta(\bold{ r}-\bold{ v}t)
\label{Lweq}
\end{equation}
(see \cite{Jackson}, section 6.3) from the wave equation (\ref{Cweq}) for
$\bold{ A}_{\rm C}$.
It is straightforward to show that the
$x$-component $A_{{\rm C}\,x}-A_{{\rm L}\,x}=(c/v)(V_{{\rm C}}-V_{{\rm L}})$
of the difference  $\bold{ A}_{\rm C}-\bold{ A}_{\rm L}$ indeed satisfies
the $x$-component of the inhomogeneous wave equation (\ref{(C-L)weq}):
\begin{equation}
\Box \left[\frac{c}{v}(V_{{\rm C}}-V_{{\rm L}})\right]
=\frac{\partial^2 V_{\rm C}}{c\partial t\partial x}.
\label{xweq}
\end{equation}
The fact that equation (\ref{xweq}) holds true follows directly from the
wave equations
\begin{equation}
\frac{c}{v}\Box V_{\rm C}=\frac{\partial^2 V_{\rm C}}{c\partial t\partial x}
-\frac{4\pi c}{v}q\delta(x-vt)\delta(y)\delta(z)
\label{boxVC}
\end{equation}
and
\begin{equation}
\Box V_{\rm L}=-4\pi q\delta(x-vt)\delta(y)\delta(z).
\label{boxVL}
\end{equation}
The wave equation (\ref{boxVC}) in turn holds true because of the facts that
the d'Alembertian $\Box=\nabla^2-\partial^2/c^2\partial t^2$,
$\nabla^2V_{\rm C}=-4\pi q\delta(x{-}vt)\delta(y)\delta(z)$,
and $\partial V_{\rm C}/\partial t=-v\partial V_{\rm C}/\partial x$;
the wave equation (\ref{boxVL}) embodies the fact that the Lorenz-gauge scalar
potential $V_{\rm L}$ is the retarded solution  of the inhomogeneous wave equation
with the right-hand side $-4\pi q\delta(x{-}vt)\delta(y)\delta(z)$.

It is also straightforward to show
that the $y$-component (\ref{ACy}) and $z$-component (\ref{ACz})
of the difference $\bold{ A}_{\rm C}-\bold{ A}_{\rm L}$
satisfy the inhomogeneous wave equation (\ref{(C-L)weq})
by using the identity
\begin{equation}
\Box(fg)=g\Box f +f\Box g+2\bold{\nabla}f\bold{\cdot\nabla}g
-2\frac{\partial f}{c\partial t}\frac{\partial g}{c\partial t}
\end{equation}
and equation (\ref{xweq}) in the evaluation of the requisite derivatives;
the singularities of the functions $y(x{-}vt)/(y^2{+}z^2)$ and 
$z(x{-}vt)/(y^2{+}z^2)$ at $y=z=0$ cannot introduce any delta-function
terms in the d'Alembertians of the components ({\ref{ACy}) and (\ref{ACz})
as the latter are functions with no singularities. 

\begin{flushleft}
{\bf 4. Poisson's equation for the vector-potential difference}
\end{flushleft}
\noindent
Onoochin \cite{Ono} attempts to solve the $x$-component of the
inhomogeneous wave equation
(\ref{(C-L)weq}) for the difference $\bold{ A}_{\rm C}-\bold{ A}_{\rm L}$ directly
by reducing it to a Poisson equation, which  is a method based on
the fact that the space and time partial derivatives are not independent
when the source term is moving uniformly (see \cite{PP}, section 19-3).
In the present case, the dependence on the variables $x$ and $t$
is only through the combination $x-vt$, and thus  
the $x$-component of the difference $\bold{ A}_{\rm C}-\bold{ A}_{\rm L}$ 
can be written as $A_{{\rm C}\,x}-A_{{\rm L}\,x}= f(x-vt,y,z)$, where
the function $f(x-vt,y,z)$ satisfies an inhomogeneous wave equation
\begin{eqnarray}
\Box f(x-vt,y,z)&=&\frac{\partial^2 V_{\rm C}}{c\partial t\partial x} 
\nonumber \\
&=&-q\frac{v}{c}\,\frac{2(x-vt)^2-y^2-z^2}{[(x-vt)^2+y^2+z^2]^{5/2}}
+q\frac{4\pi v}{3c}\delta(x-vt)\delta(y)\delta(z)
\label{boxf}
\end{eqnarray}
that can be cast as a Poisson equation on the substitutions
$\partial^2/\partial x^2-\partial^2/c^2\partial t^2
=\partial^2/\gamma^2\partial x^2$ and $x-vt=\chi/\gamma$,
where $\gamma=(1-v^2/c^2)^{-1/2}$:
\begin{equation}
\left(\frac{\partial^2}{\partial \chi^2}+\frac{\partial^2}{\partial y^2}
+\frac{\partial^2}{\partial z^2}\right) f(\chi/\gamma,y,z)
=-q\frac{v}{c}\,\frac{2\chi^2/\gamma^2-y^2-z^2}{(\chi^2/\gamma^2+y^2+z^2)^{5/2}}
+q\frac{4\pi v}{3c}\gamma\delta(\chi)\delta(y)\delta(z).
\label{Pois}
\end{equation}
The delta-function term on the right-hand side of equation (\ref{boxf})
arises from the fact that 
$\partial^2V_{\rm C}/c\partial t\partial x =-(v/c)\partial^2V_{\rm C}/\partial x^2$ 
and the delta-function identity \cite{Frahm}
\begin{equation}
\frac{\partial^2}{\partial x_i\partial x_j}\frac{1}{r}
=\frac{3x_ix_j-r^2\delta_{ij}}{r^5}-\frac{4\pi}{3}\delta_{ij}\delta(\bold{ r})
\label{laplace/3}
\end{equation}
where $r=(x_1^2+x_2^2+x_3^2)^{1/2}$, $\delta_{ij}$ is the Kronecker delta
symbol and $\delta(\bold{ r})=\delta(x_1)\delta(x_2)\delta(x_3)$ is the 
three-dimensional delta function.
Strictly speaking, the first term on the right-hand side of 
equation (\ref{laplace/3}) should be understood as 
$\lim_{a\rightarrow 0}(3x_ix_j-r^2\delta_{ij})/(r^2+a^2)^{5/2}$, 
where the limit  is to be performed after 
an ${\bf R}^3$ integration with a well-behaved `test' function.   
This limit is automatically implemented when the integration  
is done in spherical coordinates and the 
integration over the angular variables is performed first 
(without any additional transformation of variables that would shift 
the origin, of course) \cite{Frahm,Ag}.

Unlike the well-known delta-function identity 
$\nabla^2 (1/r)=-4\pi\delta(\bold{ r})$, the identity (\ref{laplace/3})
is needed only relatively rarely in electromagnetism, an example being
the calculation of the fields of electric and magnetic dipoles from 
their potentials (see \cite{Frahm}; and \cite{Jackson}, equations (4.20)
and (5.64)).
The identity $\nabla^2 (1/r)=-4\pi\delta(\bold{ r})$ obviously
requires that the second-order partial derivatives 
$\partial^2 r^{-1}/\partial x_i^2$, $i=1,2,3$ 
are given by expressions like those of equation (\ref{laplace/3});
the precise form of these can be seen most easily to follow from the 
limit $a\rightarrow 0$ of
\begin{equation}
\frac{\partial^2}{\partial x_i\partial x_j}\frac{1}{\sqrt{r^2+a^2}}
=\frac{3x_ix_j-r^2\delta_{ij}}{(r^2+a^2)^{5/2}}
-\frac{a^2\delta_{ij}}{(r^2+a^2)^{5/2}}
\label{reg}
\end{equation}
as here the limit $a\rightarrow 0$ of the second term on the right-hand side 
is a representation of $-\case{4}{3}\pi\delta_{ij}\delta(\bold{ r})$.

The standard integral expression for the solution to Poisson's equation 
(\ref{Pois}) is given by
\begin{eqnarray}
f=\frac{qv}{4\pi c}\int_{-\infty}^{\infty}{\rm d}\chi'
\int_{-\infty}^{\infty}{\rm d}y'\int_{-\infty}^{\infty}{\rm d}z'&&\left[
\frac{2\chi'^2/\gamma^2-y'^2-z'^2}{(\chi'^2/\gamma^2+y'^2+z'^2)^{5/2}}
-\frac{4\pi}{3}\gamma\delta(\chi')\delta(y')\delta(z')\right]\nonumber\\
&&\times\frac{1}{\sqrt{(\chi{-}\chi')^2+(y{-}y')^2+(z{-}z')^2}}
\label{standard}
\end{eqnarray}
where one recovers the original variables by putting $\chi=\gamma(x-vt)$. 
Here, the delta-function term is readily integrated to yield
\begin{eqnarray}
f&=&\frac{qv}{4\pi c}\int_{-\infty}^{\infty}{\rm d}\chi'
\int_{-\infty}^{\infty}{\rm d}y'\int_{-\infty}^{\infty}{\rm d}z'
\frac{2\chi'^2/\gamma^2-y'^2-z'^2}{(\chi'^2/\gamma^2+y'^2+z'^2)^{5/2}}\,
\frac{1}{\sqrt{(\chi{-}\chi')^2+(y{-}y')^2+(z{-}z')^2}}\nonumber\\
&&-\frac{qv\gamma}{3c}\frac{1}{\sqrt{\chi^2+y^2+z^2}}.
\label{standard2}
\end{eqnarray}
A direct evaluation in closed form of the three-dimensional integral in 
(\ref{standard2}) does not seem possible. However, this integral can be 
evaluated for the special case $y=z=0$, and the result is (see Appendix):
\begin{eqnarray}
\frac{qv}{4\pi c}\int_{-\infty}^{\infty}{\rm d}\chi'
\int_{-\infty}^{\infty}{\rm d}y'\int_{-\infty}^{\infty}{\rm d}z'
\frac{2\chi'^2/\gamma^2-y'^2-z'^2}{(\chi'^2/\gamma^2+y'^2+z'^2)^{5/2}}\,
\frac{1}{\sqrt{(\chi{-}\chi')^2+{y'}^2+{z'}^2}}&&\nonumber \\
=\frac{qv\gamma}{3c|\chi|}=\frac{qv}{3c|x-vt|}&&.
\label{f}
\end{eqnarray}
For $y=z=0$, the value of the second term on the right-hand side of 
(\ref{standard2}) is exactly equal and opposite to the value of (\ref{f}),
and thus the solution (\ref{standard2}) of the Poisson equation (\ref{Pois})
vanishes at $y=z=0$, as required by the result (\ref{ACx}) that we obtained
for the difference $A_{{\rm C}\,x}-A_{{\rm L}\,x}$.

Onoochin's calculation omits the delta-function term 
that arises from the second-order partial derivative of the 
Coulomb-gauge scalar potential $V_{\rm C}$, and he takes erroneously
just the first term of (\ref{standard2}) as the solution of the
Poisson equation (\ref{Pois}). 
According to equation (\ref{f}), this incorrect solution is non-zero at $y=z=0$,
and, moreover, its partial time derivative does not vanish there.
As $\partial(V_{\rm C}-V_{\rm L})/\partial x=0$ at $y=z=0$, 
the result (\ref{f}), if it were the true difference $A_{{\rm C}\,x}-A_{{\rm L}\,x}$
at $y=z=0$, would lead there to a 
non-zero difference between the Coulomb- and Lorenz-gauge $x$-components 
of the electric field. This is Onoochin's evidence against the equivalence 
of the Lorenz and Coulomb gauges.
(In his paper \cite{Ono},
Onoochin does not evaluate the integral in equation (\ref{standard2}),
but he has communicated to the present author a calculation   
of its value at $y{=}z{=}0$ that differs from the value given by 
equation (\ref{f}) by a factor of 3 \cite{perscom}.)
  
It is not difficult to show directly that 
\begin{equation}
f=\frac{c}{v}\left(\frac{q}{\sqrt{\chi^2/\gamma^2+y^2+z^2}}
-\frac{q\gamma}{\sqrt{\chi^2+y^2+z^2}}\right)
\label{fP}
\end{equation}
is the vanishing-at-infinity solution of the Poisson equation (\ref{Pois}). 
Using the relation (\ref{laplace/3}) and the differentiation rule
$\partial^2g(ax,y,z)/\partial x^2=a^2\partial^2g(u,y,z)/\partial u^2|_{u=ax}$, 
we have
\begin{eqnarray}
&&\frac{\partial^2}{\partial\chi^2}\frac{1}{\sqrt{\chi^2/\gamma^2+y^2+z^2}}
=\frac{1}{\gamma^2}\left[\frac{2\chi^2/\gamma^2-y^2-z^2}
 {(\chi^2/\gamma^2+y^2+z^2)^{5/2}}-\frac{4\pi}{3}
\gamma\delta(\chi)\delta(y)\delta(z)\right]\\
&&\frac{\partial^2}{\partial y^2}\frac{1}{\sqrt{\chi^2/\gamma^2+y^2+z^2}}
=\frac{2y^2-\chi^2/\gamma^2-z^2}
 {(\chi^2/\gamma^2+y^2+z^2)^{5/2}}-\frac{4\pi}{3}
\gamma\delta(\chi)\delta(y)\delta(z)\\
&&\frac{\partial^2}{\partial z^2}\frac{1}{\sqrt{\chi^2/\gamma^2+y^2+z^2}}
=\frac{2z^2-\chi^2/\gamma^2-y^2}
 {(\chi^2/\gamma^2+y^2+z^2)^{5/2}}-\frac{4\pi}{3}
\gamma\delta(\chi)\delta(y)\delta(z)
\end{eqnarray}
which gives
\begin{eqnarray}
&&\left(\frac{\partial^2}{\partial \chi^2} +\frac{\partial^2}{\partial y^2}
+\frac{\partial^2}{\partial z^2}\right)
\frac{1}{\sqrt{\chi^2/\gamma^2+y^2+z^2}}\nonumber \\
&&=-\left(1-\frac{1}{\gamma^2}\right)\frac{2\chi^2/\gamma^2-y^2-z^2}
{(\chi^2/\gamma^2+y^2+z^2)^{5/2}}-\frac{4\pi}{3}
\left(2+\frac{1}{\gamma^2}\right)\gamma\delta(\chi)\delta(y)\delta(z).
\label{delta1}
\end{eqnarray}
Using the relation $\nabla^2(1/r)=-4\pi\delta(\bold{ r})$, we have also
\begin{equation}
\left(\frac{\partial^2}{\partial \chi^2} +\frac{\partial^2}{\partial y^2}
+\frac{\partial^2}{\partial z^2}\right)\frac{\gamma}{\sqrt{\chi^2+y^2+z^2}}
=-4\pi\gamma\delta(\chi)\delta(y)\delta(z).      
\label{delta2}
\end{equation}
Subtracting equation (\ref{delta2}) from equation (\ref{delta1}), multiplying
the result by $qc/v$ and using the fact that $1-\gamma^{-2}=v^2/c^2$, 
we obtain
\begin{eqnarray}
\left(\frac{\partial^2}{\partial \chi^2} +\frac{\partial^2}{\partial y^2}
+\frac{\partial^2}{\partial z^2}\right)\frac{c}{v}
\left(\frac{q}{\sqrt{\chi^2/\gamma^2+y^2+z^2}}
-\frac{q\gamma}{\sqrt{\chi^2+y^2+z^2}}\right)\nonumber \\
=-q\frac{v}{c}\frac{2\chi^2/\gamma^2-y^2-z^2}
{(\chi^2/\gamma^2+y^2+z^2)^{5/2}}
+q\frac{4\pi v}{3c}\gamma\delta(\chi)\delta(y)\delta(z)
\label{regPois}
\end{eqnarray}
which shows that the function $f$ of equation (\ref{fP}) indeed satisfies the 
Poisson equation (\ref{Pois}).
Transforming back to the original variables through $\chi=\gamma(x-vt)$,
the function $f$  becomes
\begin{equation}
f=\frac{c}{v}\left[\frac{q}{\sqrt{(x-vt)^2+y^2+z^2}}
-\frac{q}{\sqrt{(x-vt)^2+\frac{1}{\gamma^2}(y^2+z^2)}}\right]
=\frac{c}{v}(V_{\rm C}-V_{\rm L})
\label{fweq}
\end{equation}
which is the value (\ref{ACx}) for the difference $A_{{\rm C}\,x}-A_{{\rm L}\,x}$.

The function $f$ of equation (\ref{fP}) has to equal
the integral solution (\ref{standard2}),
yielding for the integral in (\ref{standard2}) a closed-form expression 
\begin{eqnarray}
\frac{1}{4\pi}\int_{-\infty}^{\infty}{\rm d}\chi'
\int_{-\infty}^{\infty}{\rm d}y'\int_{-\infty}^{\infty}{\rm d}z'
\frac{2\chi'^2/\gamma^2-y'^2-z'^2}{(\chi'^2/\gamma^2+y'^2+z'^2)^{5/2}}\,
\frac{1}{\sqrt{(\chi{-}\chi')^2+(y{-}y')^2+(z{-}z')^2}}&&\nonumber \\
=\frac{c^2}{v^2}\left[\frac{1}{\sqrt{\chi^2/\gamma^2+y^2+z^2}}
-\left(1-\frac{v^2}{3c^2}\right)\frac{\gamma}{\sqrt{\chi^2+y^2+z^2}}\right]&&.
\label{ffull}
\end{eqnarray}
Direct numerical three-dimensional quadrature of
this integral, with the first factor of the integrand regularized as
mentioned above in connection with equation (\ref{laplace/3}),
resulted in values that were close to those of the
closed-form expression on the right-hand side of (\ref{ffull}).

\begin{flushleft} 
{\bf 5. Concluding remarks}
\end{flushleft}
\noindent
We found an explicit expression for the gauge function of the transformation
between the Lorenz and Coulomb gauges for a uniformly moving point charge.
The Coulomb-gauge potentials obtained using the gauge function are
guaranteed to yield the same electric and magnetic fields as the
well-known Lorenz-gauge potentials of the charge.
The expression obtained for the difference between the vector potentials in
the two gauges satisfies the Poisson equation to which the inhomogeneous
wave equation for this difference reduces after a transformation
of the variables.
However, the right-hand side of the Poisson equation involves a
delta-function term which has to be included in the integral 
expression for its solution. 

Although gauge invariance is a foregone conclusion in a gauge-invariant
theory, we believe that an explicit demonstration of the equivalence
of the Lorenz and Coulomb gauges in the basic problem of a uniformly
moving charge was instructive.

\begin{flushleft}
{\bf  Acknowledgment}
\end{flushleft}
\noindent
The author acknowledges correspondence with V V Onoochin, whose
determined objections against `mainstream' electrodynamics
motivated the writing of this paper.

\begin{flushleft}
{\bf Appendix}
\end{flushleft}
\noindent
After the transformation $\chi'/\gamma=x'$, the integral (\ref{f}) can be
written as $(qv\gamma/c)I(\gamma,X)$, where
$I(\gamma,X)$ is the integral (we drop the primes on the integration
variables):
\begin{equation}
I(\gamma,X)=\frac{1}{4\pi}
\int_{-\infty}^{\infty}{\rm d}x\int_{-\infty}^{\infty}{\rm d}y
\int_{-\infty}^{\infty}{\rm d}z\,
\frac{2x^2-y^2-z^2}{(x^2+y^2+z^2)^{5/2}}
\frac{1}{\sqrt{\gamma^2(X-x)^2+y^2+z^2}}
\label{I}
\end{equation}
where $X=\chi/\gamma$ and $\gamma>1$ are real parameters.

Transforming from the Cartesian coordinates $x,y,z$ to the spherical ones,
$r,\theta,\phi$, we have
\begin{eqnarray}
I(\gamma,X)
&=&\frac{1}{2}\int_{0}^{\infty}r^2\,{\rm d}r\int_{0}^{\pi}
\sin\theta\,{\rm d}\theta\,
\frac{r^2(3\cos^2\theta-1)}{r^5}
\frac{1}{\sqrt{\gamma^2(X-r\cos\theta)^2+r^2(1-\cos^2\theta)}}
\nonumber \\
&=&\frac{1}{2}\int_{0}^{\infty}{\rm d}r\int_{-1}^{1}{\rm d}\xi\,
\frac{3\xi^2-1}{r}\frac{1}{\sqrt{\gamma^2(X-r\xi)^2+r^2(1-\xi^2)}}.
\end{eqnarray}
We integrate with respect to $\xi$ first (this is needed
to implement the limit $a\rightarrow 0$ of the first term on the
right-hand side of equation (\ref{reg}), see \cite{Frahm,Ag}):
\begin{eqnarray}
F(r,\gamma,X)
&=&\int_{-1}^{1}{\rm d}\xi\,
\frac{3\xi^2-1}{r}\frac{1}{\sqrt{\gamma^2(X-r\xi)^2+r^2(1-\xi^2)}}
\nonumber \\
&=&\frac{1}{2\omega^5r^4}\left[3\gamma\omega(A{+}B)
+(1{+}2\gamma^2)(\omega^2r^2{-}3\gamma^2X^2)
\ln\frac{\gamma\omega|X{+}r|{-}\omega^2r{-}\gamma^2X}
{\gamma\omega|X{-}r|{+}\omega^2r{-}\gamma^2X}\right]
\label{F}
\end{eqnarray}
where
\begin{equation}
\omega=\sqrt{\gamma^2-1}\;\;\;\;\;\;
A=(\omega^2r-3\gamma^2X)|X+r|\;\;\;\;\;\;
B=(\omega^2r+3\gamma^2X)|X-r|.
\end{equation}
The function $F(r,\gamma,X)$ has the properties
\begin{equation}
F(r,\gamma,X)=F(r,\gamma,-X)\;\;\;\;\;\;
\lim_{r \rightarrow 0+}F(r,\gamma,X)=0.
\end{equation}
It peaks at $r=|X|$, where its derivative with respect to $r$
is discontinuous. The argument of the logarithm in (\ref{F})
reduces to $(\gamma|X|-\omega r)/(\gamma|X| +\omega r)$ for $r<|X|$,
and to $(\gamma-\omega)/(\gamma+\omega)=(\gamma-\omega)^2$ for $r>|X|$.

The integration with respect to $r$ is performed in two parts:
\begin{eqnarray}
I_1(\gamma,X)&=&\frac{1}{2}\int_0^{|X|}{\rm d}r\,F(r,\gamma,X)
=\frac{(2+\gamma^2)(1+2\gamma^2)}{6\gamma\omega^4|X|}
+\frac{1+2\gamma^2}{2\omega^5|X|}\ln(\gamma-\omega)\\
I_2(\gamma,X)&=&\frac{1}{2}\int_{|X|}^{\infty}{\rm d}r\,F(r,\gamma,X)
=-\frac{3\gamma}{2\omega^4|X|} 
-\frac{1+2\gamma^2}{2\omega^5|X|}\ln(\gamma-\omega)
\end{eqnarray}
where again $\omega=(\gamma^2-1)^{1/2}$. The whole integral $I(\gamma,X)$
is thus
\begin{equation}
I(\gamma,X)=I_1(\gamma,X)+I_2(\gamma,X)=\frac{1}{3\gamma|X|}
\end{equation}
and the integral (\ref{f}) has the value
$(qv\gamma/c)I(\gamma,X)=qv/(3c|X|)=qv\gamma/(3c|\chi|)$.

The results of this Appendix were obtained using  the software system
{\it Mathematica} \cite{Wolf},
and were checked by performing numerical integrations.


\begin{references}
\bibitem{Jackson}Jackson J D 1999 {\it Classical Electrodynamics} 3rd edn (New York:
                 Wiley)
\bibitem{Griffiths}Griffiths D J 1999 {\it Introduction to Electrodynamics} 3rd edn
                  (Upper Saddle River, NJ: Prentice Hall)
\bibitem{PP} Panofsky W K H and Phillips M 1962 {\it Classical Electricity and
Magnetism} 2nd edn (Reading, MA: Addison-Wesley)
\bibitem{LL}Landau L D and Lifshitz E M 1976 {\it The Classical Theory of Fields}
            4th revised English edn (Oxford: Pergamon) 
\bibitem{Rohrlich} Rohrlich F 2002 Causality, the Coulomb field, and Newton's law
                  of gravitation {\it Am. J. Phys.} {\bf 70} 411--414 \\
                  Jefimenko O D 2002
    Comment on Causality, the Coulomb field, and Newton's law of gravitation,
    by F. Rohrlich {\it Am. J. Phys.} {\bf 70} 964 \\         
   Rohrlich F 2002 Reply to Comment on Causality, the Coulomb field, and 
  Newton's law of gravitation, by O. D. Jefimenko {\it Am. J. Phys.} {\bf 70} 964
\bibitem{Jack} Jackson J D 2002 From Lorenz to Coulomb and other explicit gauge 
         transformations {\it Am. J. Phys.} {\bf 70} 917--928 
\bibitem{Stew} Stewart A M 2003 Vector potential of the Coulomb gauge
              {\it Eur. J. Phys.} {\bf 24} 519--524 \\
                Hnizdo V 2004 Comment on Vector potential of the Coulomb gauge 
               {\it Eur. J. Phys.} {\bf 25} in press
               (2003 {\it arXiv.org e-print} physics/0309028)
\bibitem{JackOk}Jackson J D  and  Okun L B 2001 Historical roots of gauge 
               invariance {\it Rev. Mod. Phys.} {\bf 73} 663--680 
\bibitem{Ono} Onoochin V V 2002 On non-equivalence of Lorentz and Coulomb
     gauges within classical electrodynamics {\it Ann. Fond. L. Broglie} 
     {\bf 27} 163--183 (2001 {\it arXiv.org e-print} physics/0111017)
    
\bibitem{Lab}Labarthe J-J 1999 The vector potential of a moving charge in the
             Coulomb gauge {\it Eur. J. Phys.} {\bf 20} L31--L32              
\bibitem{Frahm} Frahm C P 1983 Some novel delta-function identities {\it Am. J. Phys.}
             {\bf 51} 826--829 
\bibitem{Ag} Aguirregabiria J M, Hern{\'a}ndes A and Rivas M 2002
             $\delta$-function converging sequencies {\it Am. J. Phys.} {\bf 70}
             180--185             
\bibitem{perscom}Onoochin V V 2003 personal communication 
\bibitem{Wolf} Wolfram S 1999 {\it The Mathematica Book} 4th ed (Champaign,
  IL: Wolfram Media)
\end{references}
\end{document}